\documentclass[pra,aps,twocolumn,superscriptaddress]{revtex4-2}
\usepackage{bm,dcolumn,amsmath,graphicx,amsfonts,amssymb,mathtools}
\usepackage{epsfig}
\usepackage{tikz}
\usepackage{xcolor}

\begin{document}

\title{High-accuracy optical clocks based on group 16-like highly charged ions}

\newcommand{\NIST}{
National Institute of Standards and Technology, Boulder, Colorado 80305, USA}

\newcommand{\CSU}{
Department of Physics, Colorado State University, Fort Collins, Colorado 80523, USA}

\newcommand{\UNSW}{
School of Physics, University of New South Wales, Sydney 2052, Australia}

\newcommand{\IU}{ Department of Physics, Faculty of Science, Islamic University of Madinah, Madina 42351, KSA}

\author{Saleh O. Allehabi}
\affiliation{\UNSW}
\affiliation{\IU}
\author{S. M. Brewer}
\affiliation{\CSU}
\author{V. A. Dzuba}
\affiliation{\UNSW}
\author{V. V. Flambaum}
\affiliation{\UNSW}
\author{K. Beloy}
\affiliation{\NIST}
		
	
\date{\today}
	
\begin{abstract}
We identify laser-accessible transitions in group 16-like highly charged ions as candidates for high-accuracy optical clocks, including S-, Se-, and Te-like systems. For this class of ions, the ground $^{3}P_{J}$ fine structure manifold exhibits irregular (nonmonotonic in $J$) energy ordering for large enough ionization degree. We consider the $|^3P_2\rangle \longleftrightarrow |^3P_0\rangle$ (ground to first-excited state) electric quadrupole transition, performing relativistic many-body calculations of several atomic properties important for optical clock development. All ions discussed are suitable for production in small-scale ion sources and lend themselves to sympathetic cooling and quantum-logic readout with singly charged ions.
\end{abstract}
	
\maketitle
	
\section{Introduction}
\begin{figure*}[tb]
\includegraphics[width=510pt]{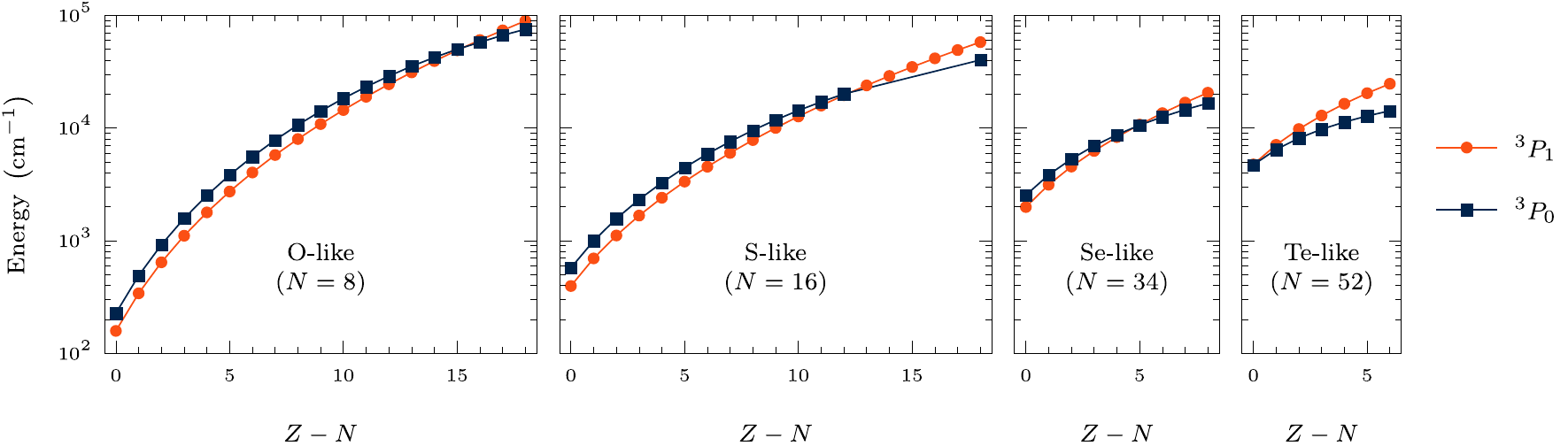}%
\caption{Experimental energies for group 16-like atomic systems. For each system, the lowest-lying electronic states are part of a $^{3}P_J$ fine structure manifold, with the $^{3}P_2$ ground state taken to have zero energy. Energies of the $^{3}P_1$ and $^{3}P_0$ states are plotted versus ionization degree $Z-N$ for the isoelectronic sequences, where $Z$ and $N$ denote the number of protons and electrons, respectively. For the O, S, and Se isoelectronic sequences, the $^{3}P_J$ energy ordering transitions from regular ordering (monotonic in $J$) at low ionization degree to irregular ordering at high ionization degree. For the Te isoelectronic sequence, the ordering is irregular already for the neutral Te system. For the systems with irregular energy ordering, the $^{3}P_0$ state lacks an $M1$ decay channel. Energies are from the NIST Atomic Spectra Database~\cite{NIST_ASD} and Refs.~\cite{Gayasov1997JPhsB,Tauheed2008CanJPhys,Reader1976JOSA}. The curves are interpolating functions intended to guide the eye.}
\label{Fig:energies}
\end{figure*}

The performance of optical clocks has improved rapidly over the last few decades~\cite{Ludlow2015RMP}.  This has led to improvements in frequency metrology as well as tests of fundamental physics using atomic clocks~\cite{Safronova2018BSM}.  The highest performance optical clocks are currently based on ensembles of neutral atoms trapped in optical lattices or singly charged ions stored in electromagnetic traps~\cite{Brewer2019PRL,McGrew2018Nature,Bothwell2019Metrologia,Huntemann2016PRL}.  However, in recent years, several clocks based on highly charged ions (HCIs) have been proposed as both improved optical frequency standards and as systems with enhanced sensitivity to possible new physics~\cite{HCI1,HCI2,HCI3} (see also Ref.~\cite{Kozlov2018RMP} and references therein).  Optical clocks based on HCIs provide several systematic advantages over current optical clocks including reduced blackbody radiation (BBR), Zeeman, and electric quadrupole shifts~\cite{HCI4,Kozlov2018RMP}. Here, we identify group 16-like HCIs as optical clock candidates. For this class of ions, the ground $^{3}P_{J}$ fine structure manifold exhibits irregular (nonmonotonic in $J$) energy ordering for large enough ionization degree, with the $^3P_1$ state lying above the $^3P_2$ (ground) and $^3P_0$ (first-excited) states. Given this irregular ordering, the $^3P_0$ excited state lacks a magnetic dipole ($M1$) decay channel, resulting in a relatively long lifetime and making the $|^3P_2\rangle \longleftrightarrow |^3P_0\rangle$ electric quadrupole ($E2$) transition a viable clock transition. This irregular energy ordering is illustrated in Fig.~\ref{Fig:energies}. Due to the high nuclear charge, the ordering is irregular for Te-like systems beginning with neutral tellurium.  In the case of O-, S-, and Se-like systems, the ionization degree must be increased before the irregular ordering is observed. Specifically, for O-like ions, the irregular ordering is not observed until Mn$^{17+}$. For this system, the clock transition wavelength ($\approx 150$~nm~\cite{NIST_ASD, CHENG1979ADNDT}) is outside the range of current clock lasers. The S-, Se-, and Te-like systems offer more favorable clock transition wavelengths. In the present work, we perform relativistic many-body calculations of relevant properties for optical clock development. While we present results only for select S-, Se-, and Te-like systems, other group 16-like systems not explicitly considered may also be of interest.

The present work is a broader study of the group 16-like systems started in our previous work with just Ba$^{4+}$~\cite{Beloy2020PRL}. We include Ba$^{4+}$ in the list of ions considered here. Broadly speaking, similar computational techniques are used here as in Ref.~\cite{Beloy2020PRL}. The results of Ref.~\cite{Beloy2020PRL} are reproduced with only small deviations, with two exceptions. First, a small clerical error is corrected, giving a second order Zeeman shift that is a factor of two larger. The essential conclusion that this shift is negligible remains valid. Second, an improved method is used to calculate the scalar differential polarizability $\Delta \alpha$, which predicts a much larger degree of cancellation between the clock state polarizabilities. While the revised value of $\Delta \alpha$ does not support cancellation between the trap-induced Stark and micromotion time-dilation shifts (by operating at a “magic” rf trap drive frequency ~\cite{Berkeland1998JAP}), it does offer highly suppressed Stark shifts, including the BBR shift. We find similar cancellation between the clock state polarzbilities for the other group 16-like systems, resulting in similarly small $\Delta \alpha$.

In the present work, we focus on the isotopes with zero nuclear spin to avoid complications caused by the hyperfine structure (hfs). In particular, the second-order Zeeman shift is enhanced in isotopes with hfs, by small hfs energy intervals. In contrast, the second-order Zeeman shift is small and can be neglected in spin-zero isotopes.	
	
	\section{\label{sec:Method}Method}
	
	\subsection{Calculation of energy levels}

The calculations are carried out using a combination of the configuration interaction (CI) technique with the linearized single-double–coupled-cluster (SD) method, as described in Ref.~\cite{Dzuba2014PRA}. The combined method (CI+SD) has been demonstrated to be efficient and very precise for systems with several valence electrons. With the SD technique, it is possible to accurately determine the core-valence and core-core electron correlations, while the CI method takes the valence-valence correlations into account. Our calculations are done using the V$^{N-M}$ approximation~\cite{Dzuba2014PRA}, where $N$ is the total number of electrons and $M$ is the number of valence electrons. For all atomic systems considered (see, e.g., Table~\ref{t:Energy}), the calculations begin with the relativistic Hartree-Fock (RHF) method for a closed-shell core, which removes all valence electrons.
We treat all systems as $M = 6$ valence systems, except for Te and Sr$^{22+}$, which are treated as $M=4$ valence systems; this is because NIST data~\cite{NIST_ASD} indicate that Te and Sr$^{22+}$ have no low-lying states with the excitations from the $5s$ and $3s$ subshells, respectively. Therefore, it is reasonable to treat $5s$ electrons in Te and $3s$ electrons in Sr$^{22+}$ as core electrons. The RHF Hamiltonian has the following form:
	\begin{equation} 
	\hat H^{\rm RHF}= c\bm{\alpha}\cdot\mathbf{p}+(\beta -1)mc^2+V_{\rm nuc}(r)+V_{\rm core}(r),
	\end{equation}
	where $c$ is the speed of light, $\bm{\alpha}$ and $\beta$ are the Dirac matrices, $\bm{p}$ is the electron momentum, $m$ is the electron mass, $V_{\rm nuc}$ is the nuclear potential obtained by integrating the Fermi distribution of the nuclear charge density, and $V_{\rm core}(r)$ is the self-consistent RHF potential created by the electrons of the closed-shell core. 
	
	Following the completion of the self-consistent procedure for the core, the B-spline technique~\cite{Johnson1986PRL,Johnson1988PRA} is used to develop a complete set of single-electron wave functions. Based on B-splines, one can make linear combinations of basis states, which are eigenstates of the RHF Hamiltonian. The basis set is built up of 40 B-splines of order 9 in a box that has a radius $R_{\rm max}=40a_B$, where $a_B$ is the Bohr radius, with the orbital angular momentum 0~$\leq$~\textit{l}~$\leq$~6. There are two types of basis states: core states and valence states. Core states are used to calculate the effective potential of the core. Valence states are used as a basis for the SD equations and for obtaining the many-electron states required for the CI calculations. 
	
	 In the process of solving the SD equations for the core and valence states, we generate correlation operators $\Sigma_1$ and $\Sigma_2$~\cite{CPM,Sigma2,Dzuba2014PRA}. $\Sigma_1$ is the correlation interaction between a particular valence electron and the core, and accordingly, one-body part $ \hat h_{1}$ can be described as follows
\begin{equation} 
	 \hat h_{1}=\hat H^{\rm RHF}+\Sigma_1 .
\end{equation}
$\Sigma_2$ represents the screening of the Coulomb interaction between a pair of valence electrons; hence, the two-body Coulomb interaction operator, $ \hat h_{2}$, is modified so as to include the two-body part of the core-valence interaction as follows (we use Gaussian electromagnetic expressions, $e$ is electron charge)
\begin{equation} 
	 \hat h_{2}=\frac{e^2}{|r_i-r_j|}+ \Sigma_{2} .
\end{equation}
Whenever there are more than one valence electron above the closed-shell core, these $\Sigma$ operators can be used in the subsequent CI calculations to account for the core-valence and core-core correlations. By solving the SD equations for external states, the single-electron energies of an atom or ion with one valence electron can also be obtained. However, we note that there are slight differences between the SD equations used for this purpose and those to be used for CI calculations. In this case, one term in the SD equations needs to be eliminated because its contribution is accounted for by the CI calculations (refer to Ref. \cite{Dzuba2014PRA}). This contribution is relatively small; therefore, differences in the SD equations can be ignored.
	 
	 In the CI approach, we build the effective CI+SD Hamiltonian for many valence electrons as a sum of one- and two-electron parts with the addition of $\Sigma_1$ and $\Sigma_2$ operators in order to account for the correlation between core and valence electrons,
	 \begin{eqnarray} \label{e:HCI}
	 \hat H^{\rm eff}=\sum_{i=1}^{M} \left(\hat H^{\rm RHF}+\Sigma_1\right)_i 
	 +\sum_{i<j}^{M} \left(\dfrac{e^2}{|r_i-r_j|}+ \Sigma_{2ij}\right)
	 \end{eqnarray}
	where $i$ and $j$ enumerate valence electrons.
	
It is well-recognized that increasing the number of valence electrons exponentially increases the size of the CI matrix. Our present work has up to six valence electrons, which leads to an extremely large CI matrix. In order to deal with a matrix of this magnitude, it would require considerable computational power. However, the size of the CI matrix can be decreased by orders of magnitude at the expense of some accuracy. In order to accomplish this, we use the recently developed version of the CI method called the CIPT method \cite{Dzuba2017PRA}. The method combines CI with perturbation theory (PT) and is used to ignore the off-diagonal matrix elements between high-energy states in the CI matrix. This step is justified because the high-energy states provide only a minimal correction to the wave function.
	
	
The wave function for valence electrons  is presented as an expansion over single-determinant basis states, which is divided in two parts,
	\begin{equation} \label{e:Wave}
	\begin{aligned}
	\Psi\left(r_{1}, \ldots, r_{M}\right)=& \sum_{i=1}^{N_{\text {eff }}} c_{i} \Phi_{i}\left(r_{1}, \ldots, r_{M}\right) \\
	&+\sum_{i=N_{\text {eff }}+1}^{N_{\text {total}}} c_{i} \Phi_{i}\left(r_{1}, \ldots, r_{M}\right) .
	\end{aligned}
	\end{equation}
Here $c_{i}$ are the expansion coefficients and $\Phi_{i}$ are single-determinant many-electron basis functions. 
The first part of the wave function represents a small number of low-energy terms that contribute a great deal to the CI valence wave function (1 $\leqslant$ $i$ $\leqslant$ $N_{\text{eff}}$, where $N_{\text{eff}}$ is the number of low-energy basis states), while the second part represents a large number of high-energy states that introduce minor corrections to the valence wave function ($N_{\text {eff }}$ $<$ $i$ $\leqslant$ $N_{\text {total}}$, where $N_{\text {total }}$is the total number of the basis states). Consequently, this allows us to truncate the CI Hamiltonian by ignoring the off-diagonal matrix elements between terms in the second summation in Eq.~(\ref{e:Wave}) $\left(\left\langle i\left|H^{\mathrm{eff}}\right| h\right\rangle=0 \text { for } N_{\text {eff }}< i, h \leqslant N_{\text {total}}\right)$, which in turn reduces computation time with a negligible loss in precision.
	
The matrix elements between low-energy states $i$ and $g$ are corrected by the following formula similar to the second-order perturbative correction to the energy	
	\begin{equation}
	\langle i|H^\mathrm{eff}|g\rangle\rightarrow\langle i|H^\mathrm{eff}|g\rangle+\sum_{k}\frac{\langle i|H^\mathrm{eff}|k\rangle\langle k|H^\mathrm{eff}|g\rangle}{E-E_{k}}.
	\label{e:CIPT}
	\end{equation}
Here, $i, g \leqslant N_{\text {eff }}, N_{\text {eff }}<k \leqslant N_{\text {total}}$, $E$ is the energy of the state of interest, and $E_k$ denotes the diagonal matrix element for high-energy states, $E_{k}=\left\langle k\left|H^{\mathrm{eff}}\right| k\right\rangle$. The summation in (\ref{e:CIPT}) runs over all high-energy states. 
Note that neglecting off-diagonal matrix elements between highly excited states corresponds to neglecting the third-order contribution
\begin{equation}
\delta E_{ig}^{(3)} =\sum_{k,l}\frac{\langle i|H^\mathrm{eff}|k\rangle \langle k|H^\mathrm{eff}|l\rangle  \langle l|H^\mathrm{eff}|g\rangle}{(E-E_{k})(E-E_{l})}.
\label{e:E3}
\end{equation}
This contribution is supressed by large energy denominators. Neglecting the third-order corrections over the second-order corrections cannot cause any false contributions to the spin-orbit splitting  or break the symmetry of the CI Hamiltonian.

The problem of finding the wave function and corresponding energy can be reduced to a modified CI matrix eigenvalue equation $\hat H^{\rm eff}$ [Eq.~(\ref{e:HCI})]  with size $N_{\text {eff }}$
	\begin{equation}
	\left( \hat H^{\rm eff}-EI\right) X = 0,
	\label{e:CI} 
	\end{equation}
where $I$ is the identity matrix and $X$ is the vector $\left\{c_{1}, \ldots, c_{N_{\text {eff }}}\right\}$.  
 Note that for accurate solution the energy parameter $E$ must be the same in Eqs.~(\ref{e:CIPT}) and (\ref{e:CI}). Since this energy is not known in advance, the equations (\ref{e:CIPT}) and (\ref{e:CI}) are solved by iterations. The starting point for the iterations can be, e.g. the solution of (\ref{e:CI}) with the matrix (\ref{e:CIPT}) without the second-order corrections.
A more comprehensive description of this technique is given in Ref.~\cite{Dzuba2017PRA}.
	
	\subsection{Calculation of transition amplitudes and lifetimes}
	
The method we use for computing transition amplitudes is based on the time-dependent Hartree-Fock (TDHF) method~\cite{Dzuba1987JPhysB}, which is the same as the well-known random phase approximation (RPA). The RPA equations are defined as
	\begin{equation}\label{e:RPA}
	\left(\hat H^{\rm RHF}-\epsilon_c\right)\delta\psi_c=-\left(\hat f+\delta V^{f}_{\rm core}\right)\psi_c,
	\end{equation}
where the operator $\hat f$ refers to an external field. The index $c$ denotes single-electron states, $\psi_c$ is a single electron wave function with corresponding energy $\epsilon_c$, $\delta\psi_c$ is a correction to the wave function due to the external field, and $\delta V^{f}_{\rm core}$ is the correction to the self-consistent RHF potential caused by the amendment of all core states in the external field. For all states in the core, the RPA equations (\ref{e:RPA}) are solved self-consistently. The transition amplitudes are found by calculating matrix elements between states $a$ and $b$ using the formula 
	\begin{equation}
	A_{a b}=\left\langle b\left|\hat f+\delta V^{f}_{\rm core }\right| a\right\rangle.
	\end{equation}
Here, $|a\rangle$ and $|b\rangle$ are the many-electron wave functions calculated with the method described above. These wave functions are given by Eq.~(\ref{e:Wave}).  In the present work, only the rates of $E2$ transitions 
are taken into account. The rates are computed as follows (in atomic units)
	\begin{equation}
	T_{ab} = \frac{1}{15}\left(\alpha\omega_{ab}\right)^5\frac{A_{ab}^2}{2J_b+1},
	\end{equation}
where $\alpha$ is the fine structure constant ($\alpha\approx\frac{1}{137}$), $\omega_{ab}$ is the frequency of the transition, $J_b$ is the total angular momentum of the upper state $b$, and $A_{ab}$ represents the transition amplitude (reduced matrix element) of the $E2$ operator. The lifetimes of each excited state $b$, $\tau_b$, expressed in seconds, are given as
	\begin{equation}
	\tau_b =  2.4189\times 10^{-17}\bigg/\sum_a T_{ab}
	\end{equation}
where the summation runs over all possible transitions to lower states $a$.
	
\section{\label{sec:Results}Results}

\subsection{Energy levels, transition amplitudes, and lifetimes of the systems}

Table \ref{t:Energy} presents the calculated energy levels of the systems and compares them to the results of previous work; note that all earlier data presented in the table are either experimental or semi-empirical, except for the value for Cd$^{14+}$, which has been calculated.
The calculated energies are in good agreement with experiment, within a few percent.
In Table~\ref{t:Energy}, we also present the $E2$ amplitudes and corresponding decay rates for excited clock states decaying to the ground state.
The rates are in good agreement with previous studies. The rates and lifetimes of the excited clock states were calculated using calculated amplitudes and experimental energies.
	
		\begin{table*}
		
		\caption{\label{t:Energy}
			Excitation energies ($E$), wavelength transitions ( $\lambda$ ), $E2$ amplitudes ($A$), decay rates ($T$), and lifetimes ($\tau$) for the excited clock states. Note that for calculating $\lambda$, the experimental energies (where available) have been used.}
		\begin{ruledtabular}
			\begin{tabular}{lrrrr c ccc}
				&&
				\multicolumn{2}{c}{$E$ [cm$^{-1}$]}&
							\multicolumn{1}{c}{$\lambda$ [nm]}&
			\multicolumn{1}{c}{$A$ [a.u]}&
\multicolumn{2}{c}{$T$ [s$^{-1}$]}&

				\multicolumn{1}{c}{$\tau$ [s]}\\
				
				\cline{3-4}
			
				\cline{7-8}
				
				\multicolumn{1}{c}{System}& 
				\multicolumn{1}{c}{State}&

				\multicolumn{1}{c}{Present }&
				\multicolumn{1}{l}{Other}&
								\multicolumn{1}{c}{}&
				\multicolumn{1}{c}{Present}&
				
				\multicolumn{1}{c}{Present }&
				\multicolumn{1}{c}{Other cal.}&
				\multicolumn{1}{c}{Present}\\
				
				\hline
				\multicolumn{9}{c}{\textbf{Te-like systems}}\\

Te & $5p^{4}$~$^3${P}$_{0}$&4630&4706\footnotemark[1]&2124.9&$-5.483$ &0.0078&0.0073\footnotemark[6], 0.0097\footnotemark[7]&128.21\\
Xe$^{2+}$ &$5s^{2}5p^{4}$~$^3${P}$_{0}$&8515&8130\footnotemark[1]&				1230.0&$-3.163$&0.0398&0.04451\footnotemark[7]&25.13\\
Ba$^{4+}$ &$5s^{2}5p^{4}$~$^3${P}$_{0}$&11548&11302\footnotemark[1]&				884.8&$-2.351$&0.1141&0.1253\footnotemark[7]&8.76\\
Ce$^{6+}$ &$5s^{2}5p^{4}$~$^3${P}$_{0}$&14697&14210\footnotemark[3]&				703.7&$-1.825$&0.2159&0.2437\footnotemark[7]&4.63\\

\multicolumn{9}{c}{\textbf{Se-like systems}}\\

Zr$^{6+}$ &$4s^{2}4p^{4}$~$^3${P}$_{0}$&12722&12557\footnotemark[4]&				796.4&$-1.131$&0.0447&0.0468\footnotemark[8]&22.37\\
Cd$^{14+}$ &$4s^{2}4p^{4}$~$^3${P}$_{0}$&28909&28828\footnotemark[5]&				345.9&0.585&0.7612&$-$&1.31\\

\multicolumn{9}{c}{\textbf{S-like systems}}\\

Ge$^{16+}$ &$3s^{2}3p^{4}$~$^3${P}$_{0}$&33635&33290\footnotemark[1]&				300.4&0.228&0.2377&0.2502\footnotemark[8]&4.21\\
Kr$^{20+}$ &$3s^{2}3p^{4}$~$^3${P}$_{0}$&47618&46900\footnotemark[1]&				213.2&0.176&0.7859&0.8322\footnotemark[8]&1.27\\
Sr$^{22+}$ &$3p^{4}$~$^3${P}$_{0}$&50911&53400\footnotemark[1]&				187.3&$-0.160$&1.2434&1.257\footnotemark[8]&0.805\\

\end{tabular}
\footnotetext[1] {  Ref.~\cite{NIST_ASD}; The values are compiled from the NIST database; Te-like systems [Expt.], S-like systems [Expt. or Semi.].}
\footnotetext[2] { Ref.~\cite{Gayasov1997JPhsB};  Expt.}
\footnotetext[3] { Ref.~\cite{Tauheed2008CanJPhys}; Expt.}
\footnotetext[4] { Ref.~\cite{Reader1976JOSA};  Expt.}
\footnotetext[5] { Ref.~\cite{Wang2017ADNDT};  Theor.}
			
\footnotetext[6] { Ref.~\cite{Garstand1964JRNBSAPC}; Theor.}
\footnotetext[7] { Ref.~\cite{Biemont1995AASS};  Theor.}

\footnotetext[8] { Ref.~\cite{Biemont1986PhysScr};  Theor.}
			
\end{ruledtabular}
	
\end{table*}

\begin{table*}
		
		\caption{\label{t:IP}
		Ionization potential (IP, cm$^{-1}$), quadrupole moment ($\Theta$, a.u.), and Landé \textit{g}-factor of the ground state.} 
		\begin{ruledtabular}
			\begin{tabular}{lrcccl}
				&&
			    	\multicolumn{2}{c}{IP}\\
					\cline{3-4}

							\multicolumn{1}{c}{System}& 
					\multicolumn{1}{c}{State}&
				
	\multicolumn{1}{c}{Present }&
\multicolumn{1}{c}{NIST}&
\multicolumn{1}{c}{$\Theta$}&
\multicolumn{1}{c}{ \textit{g}-factor}\\

				\hline
				
					\multicolumn{6}{c}{\textbf{Te-like systems}}\\
					
				Te & $5p^{4}$~$^3${P}$_{2}$&70939&72669.006(0.047)&1.22&1.467\footnotemark[1]\\
				Xe$^{2+}$ &$5s^{2}5p^{4}$~$^3${P}$_{2}$&247505&250400(300)&0.53&1.441\\
				Ba$^{4+}$ &$5s^{2}5p^{4}$~$^3${P}$_{2}$&475734&468000(15000)&0.32&1.424\footnotemark[2]\\
				Ce$^{6+}$ &$5s^{2}5p^{4}$~$^3${P}$_{2}$&748287&734000(16000)&0.19&1.407\\
				
				\multicolumn{6}{c}{\textbf{Se-like systems}}\\
				
				Zr$^{6+}$ &$4s^{2}4p^{4}$~$^3${P}$_{2}$&913400&903000(16000)&0.23&1.457\\
				Cd$^{14+}$ &$4s^{2}4p^{4}$~$^3${P}$_{2}$&2894809&2887000(22000)&0.062&1.407\\

				\multicolumn{6}{c}{\textbf{S-like systems}}\\
				Ge$^{16+}$ &$3s^{2}3p^{4}$~$^3${P}$_{2}$&4916400&4912400(6800)&0.042&1.445\\
				Kr$^{20+}$ &$3s^{2}3p^{4}$~$^3${P}$_{2}$&7126586&7120300(10100)&0.024&1.420\\
				
				Sr$^{22+}$ &$3p^{4}$~$^3${P}$_{2}$&8378219&8372100(12000)&0.019&1.413\\

			\end{tabular}

		\end{ruledtabular}
		
			\footnotetext[1]{Experimental value is 1.460(4)~\cite{NIST_ASD}.}
			\footnotetext[2]{The same as in our previous calculations, 1.42~\cite{Beloy2020PRL}.}
\end{table*}

\begin{table*}
		\caption{\label{t:pol}
			Scalar static polarizabilities of the ground and excited clock states ($\alpha_0({\rm GS})$ and $\alpha_0({\rm ES})$, respectively), and BBR frequency shifts for the clock transition.  $\delta\nu_{\rm BBR}$/$\nu$ is the fractional contribution of the BBR shift, where $\nu$ is the clock transition frequency. ``Total'' means total scalar polarizability (core + valence). 
Error bars were obtained on the assumption that the accuracy for the polarizability is 10\%. The last column shows the second-order Zeeman shifts, $\delta\nu_{\rm SZ}$. The notation $x[y]$ abbreviates $x\times10^y$.}
		\begin{ruledtabular}
			\begin{tabular}{lcccccllllc}
				&
				\multicolumn{1}{c}{$\alpha_0$~[a.u.]}&
				\multicolumn{2}{c}{$\alpha_0({\rm GS})$~[a.u.]}&
				\multicolumn{2}{c}{$\alpha_0({\rm ES})$~[a.u.]}&&
				\multicolumn{3}{c}{BBR (\textit{T} = 300 K)} &
					\multicolumn{1}{c}{$\delta\nu_{\rm SZ}$}\\
			
				\cline{3-4}
				\cline{5-6}
				\cline{8-10}
				\multicolumn{1}{c}{System}&

				\multicolumn{1}{c}{ Core}&
				
				\multicolumn{1}{c}{Valence} &
				\multicolumn{1}{c}{Total} &

				\multicolumn{1}{c}{Valence} &
				\multicolumn{1}{c}{Total} &
				\multicolumn{1}{c}{$\Delta \alpha (0)$} &
				\multicolumn{1}{c}{$\delta\nu_{\rm BBR}$~[Hz]}&
				\multicolumn{1}{c}{$\nu$~[Hz]}&
				\multicolumn{1}{c}{$\delta\nu_{\rm BBR}$/$\nu$} &
					\multicolumn{1}{c}{[Hz/(mT)$^2]$}\\
				\hline
				
		\multicolumn{11}{c}{\textbf{Te-like systems}}\\
				
Te &8.84& 28.5 &37.3\footnotemark[1]& 29.6& 38.4&$<7$&$<6[-2]$&$1.411[14]$&$<4[-16]$ &$-87$\\
				
Xe$^{2+}$ &0.835&10.1&10.9&10.4&11.2&$<2$&$<2[-2]$&$2.437[14]$&$<7[-17]$&$-2.02$\\
				
Ba$^{4+}$&0.578 &5.46&6.04&5.56&6.14&$<1$&$<1[-2]$&$3.388[14]$&$<3[-17]$&$-0.55$\\																

Ce$^{6+}$&0.421 &3.43&3.85&3.49&3.91&$<0.6$&$<5[-3]$&$4.260[14]$&$<1[-17]$&$-0.22$\\

	\multicolumn{11}{c}{\textbf{Se-like systems}}\\

Zr$^{6+}$ &0.083&1.87&1.95&1.89&1.97&$<0.1$&$<1[-3]$&$3.764[14]$&$<1[-18]$&$-3.72$\\
Cd$^{14+}$ &0.024&0.466&0.490&0.467&0.491&$<1[-2]$&$<1[-5]$&$8.642[14]$&$<1[-20]$&$-0.08$\\

	\multicolumn{11}{c}{\textbf{S-like systems}}\\
	
					Ge$^{16+}$ &0.002&0.142&0.144&0.142&0.144&$<1[-3]$&$<1[-5]$&$9.980[14]$&
					$<1[-20]$&$-0.39$\\
	Kr$^{20+}$ &0.001&$0.0812$&$0.0822$&$0.0810$&$0.0820$&$<1[-3]$&$<1[-5]$&$1.406[15]$&
	$<1[-20]$&$-0.08$\\
	
	Sr$^{22+}$ &0.063&$0.0276$&$0.0906$&$0.0273$&$0.0903$&$<1[-3]$&$<1[-5]$&$1.601[15]$&
	$<1[-20]$&$-0.04$\\
					
\end{tabular}
\footnotetext[1]{The polarizability of the Te atom has been studied before, and the recommended result is 38$\pm$4 a.u.~\cite{Schwerdtfegera2019MolPhys}.}
		\end{ruledtabular}
		
	\end{table*}

\subsection{Ionization potential, Land\'{e} $g$-factors, and electric quadrupole moments}

Table~\ref{t:IP} presents the results of the calculated ionization potential (IP) of all atomic systems. The ionization potential (IP) of a system can be calculated as a difference in the ground state energy between the system of interest ($E^M$) and the following ion ($E^{M-1}$),  IP = $E^{M-1} - E^M$.  The results of our calculations are compared with data compiled by NIST. With the exception of the first two systems, the NIST data have large uncertainties ranging from 6800 cm$^{-1}$ to 22000 cm$^{-1}$. Within these uncertainties, our calculations agree with the NIST data.
In Table~\ref{t:IP}, we also present the calculated values of the Land\'{e} $g$-factors for the ground states of all systems. The $g$-factors are calculated as expectation values of the $M1$ operator.
 



Electric quadrupole shifts are known to be caused by an interaction between the quadrupole moment of an atomic state and an external electric-field gradient, and in the Hamiltonian, the corresponding term is given as \cite{Itano2000NISTJRes}
\begin{equation}
H_{\mathcal{Q}}=\sum_{q=1}^{-1}(-1)^{q} \nabla \mathcal{E}_{q}^{(2)} \hat \Theta_{-q}.
\end{equation}
Here, the tensor $ \nabla \mathcal{E}_{q}^{(2)}$ represents the external electric field gradient at the atom's position, and $\hat \Theta_{q}$ describes the electric-quadrupole operator for the atom. It is the same as for the $E2$ transitions, $\hat \Theta_q = r^2C^{(2)}_q$, where $C^{(2)}_q$ is the normalized spherical function and
$q$ indicates the operator component.  
The electric quadrupole moment, $\Theta$, is defined as the expectation value of $\hat\Theta_{0}$ for the extended state
\begin{equation}
\begin{aligned}
\Theta &=\left\langle n J J\left|\hat \Theta_{0}\right| n J J\right\rangle\\
&=\left\langle n J\|\hat\Theta\| n J\right\rangle \sqrt{\frac{J\left(2 J-1\right)}{\left(2 J+3\right)\left(2 J+1\right)\left(J+1\right)}},
\end{aligned}
\end{equation}
where $\left\langle n J\|\hat\Theta\| n J\right\rangle$ indicates the reduced matrix element of the electric quadrupole operator. We compute the values of $\Theta$ using the CI+SD and RPA methods described in the previous section. The results are presented in Table \ref{t:IP}. Note that the excited clock states of all atomic systems have $\Theta = 0$ since the total angular momentum $J$ is zero. Some of these atomic systems have been investigated before.
In our early work~\cite{Kozlov2014PRA} a different approach was used leading to quadrupole moments $Q({\rm Te})=-2.58$~a.u. and $Q({\rm Xe}^{2+})=-1.17$~a.u.
It should be noted that in this earlier work~\cite{Kozlov2014PRA}, the electric quadrupole moment $Q$ is defined in a way which differs from our definition by a factor of 2, so that $\Theta$ = $Q/2$. Taking this into account, the results for the two calculations are in good agreement. 
	
	 
\subsection{Polarizabilities, blackbody radiation shifts, and  second-order Zeeman shifts}

The scalar polarizability $\alpha_v(0)$ of an atomic system in state $v$ is given by a sum over a complete set of excited states $n$ connected to state $v$ by 
the electric dipole ($E1$) reduced matrix elements (we use atomic units) 
\begin{equation}\label{e:pol}
\alpha_v(0)=\dfrac{2}{3(2J_v+1)}\sum_{n}\frac{A_{vn}^2}{\omega_{vn}},
\end{equation}
where $J_v$ is the total angular momentum of state $v$ and $\omega_{vn}$ is the frequency of the transition. Notations $v$ and $n$ refer to many-electron atomic states. For the calculations of the polarizabilities of clock states, we apply the technique developed in Ref.~\cite{Dzuba2020Symmetry} for atoms or ions with open shells. The method relies on Eq.~(\ref{e:pol}) and the Dalgarno-Lewis approach~\cite{Dalgarno1955PRS}, which reduces the summation in Eq.~(\ref{e:pol}) to
solving a matrix equation (see Ref.~\cite{Dzuba2020Symmetry} for more details).

Results for the polarizabilities of the ground and excited clock states are shown in Table \ref{t:pol}. 
It appears that the polarizabilities of the ground and excited clock states of all atomic systems are similar in values. This is because both clock states belong to the same fine structure manifold, and the energy intervals between them is significantly smaller than the excitation energies to the opposite-parity states (see Eq.~(\ref{e:pol})).
	
Some of these atomic systems have previously been studied for their polarizabilities. Review~\cite{Schwerdtfegera2019MolPhys} and references therein have investigated the ground state polarizability of Te both theoretically and experimentally, and the recommended value has been determined to be 38$\pm$4 a.u. Compared with the recommended value, our calculation (37.3~a.u.)\ is in excellent agreement.
In our earlier work \cite{Kozlov2014PRA} a simplified approach were used leading to larger values of polarizabities of Te and Xe$^{2+}$; 45.96~a.u. and 47.80~a.u. for lower and upper clock states of Te, and 14.69~a.u. and  14.79~a.u. for lower and upper clock states of Xe$^{2+}$.
These results are in reasonable agreement with our present calculations.

In our previous work~\cite{Beloy2020PRL}, we calculated the polarizability of the ground and excited clock states for Ba$^{4+}$ and found the values to be 4.4 a.u and 1.4 a.u., respectively. Those results are in disagreement with the present results. The reason for the disagreement comes from the fact that direct summation was used in Ref.~\cite{Beloy2020PRL}. This method works well if the summation is strongly dominated by the contribution of the low-lying states of opposite parity. This is not the case for Ba$^{4+}$ or the other systems considered here. In this paper, we use the more accurate method described above.
The accuracy of the current approach can be judged by recalling our earlier calculations~\cite{EDM6,EDM-Tl,DD-Yb10}. Deviation of the calculated polarizabilities from the experimental values varies from fraction of percent for noble elements~\cite{EDM6} to few percent for atoms with more complicated electron structure.  Given also that we have excellent agreement for Te with the recommended value from literature, which has 10\% uncertainty, we conclude that the accuracy of our present calculations is in the range from 1\% to 10\%.

Blackbody radiation (BBR) can have a significant impact on the clock transition frequency in atomic clocks.
The shift in the clock transition frequency caused by BBR can be calculated as
\begin{equation}
\delta \nu_{\rm BBR} = -1.063 \times  10^{-12}T^4 \Delta \alpha(0),
\end{equation}
where $T$ is the temperature and $ \Delta \alpha(0)= \alpha_0({\rm ES}) - \alpha_0({\rm GS})$ is the difference between the excited and ground clock-state polarizabilities. The proportionality factor here is for a shift in Hz, temperature in K, and differential polarizability in atomic units. The results of the fractional BBR shifts at room temperature are shown in Table~\ref{t:pol}. It can be seen from the table that the differential polarizabilities are extremely small, which results in small values for BBR shifts.
Note that even the use of the most optimistic assumption about the accuracy of the calculation of the polarizabilities (1\%) leads to large uncertainties in the BBR shift. This means that the numbers for the BBR shift in Table~\ref{t:pol} should be considered as upper limits.

In order to calculate the second-order Zeeman shift ($\delta\nu_{\rm SZ}$), we have to take into account an influence caused by a weak homogeneous external magnetic field. For the determination of $\delta\nu_{\rm SZ}$, the following formula can be used \cite{Porsev2020PRA}
\begin{equation}
\delta\nu_{\rm SZ}=-\frac{1}{2h} \Delta\alpha^{\mathrm{M} 1} B^{2},
\end{equation} 
where $h$ is Planck's constant, $B$ is the magnetic field, and $\Delta\alpha^{\mathrm{M} 1}$ is the difference between the magnetic-dipole polarizability of the ground and excited clock states, $\Delta\alpha^{\mathrm{M} 1}= \alpha^{\mathrm{M} 1}(\rm ES)-\alpha^{\mathrm{M} 1}(\rm GS)$. The $M1$ polarizability can be calculated using Eq.~(\ref{e:pol}), but the amplitude of the electric dipole transitions ($A_{vn}$) should be replaced with the amplitude of the magnetic dipole transitions. Our results are shown in Table \ref{t:pol}. It should be mentioned that the magnetic-dipole polarizabilities can be calculated with just a few low-lying states since their contributions dominate. In the case of the atomic systems consider here, only the first two low-lying states belonging to the same configuration give significant contributions. 
Here only the scalar contribution is presented. A tensor contribution of similar magnitude also exists, though it can be canceled with certain averaging schemes. In any case, the scalar results illustrate the scale of the second order Zeeman shift, which is negligibly small for small ($\sim\!\mu$T) magnetic fields.

\subsection{Sensitivity of the clock transitions to variation of the fine structure constant}

Variations in the fine structure constant could lead to an observable effect on the clock transition frequency. The relationship between the clock frequency and the fine-structure constant in the vicinity of their physical values can be expressed as:
\begin{equation}\label{e:q}
\omega = \omega_0 + q\left[\left(\frac{\alpha}{\alpha_0}\right)^2-1\right],
\end{equation}
where $\alpha_0$ and $\omega_0$ are the laboratory values of the fine structure constant and the transition frequency, respectively, and $q$ is the sensitivity coefficient that is determined from atomic calculations \cite{Flambaum2009CanJPhys}. 
Note that we do not consider variation of atomic unit of energy $m_e e^4/\hbar^2$ since it cancels out in the ratio of frequencies. Variation of dimensionfull parameters  like  $m_e e^4/\hbar^2$  depend on the units one uses. For example, in atomic units it is equal to one and does not vary. Therefore, dependence of frequencies on $\alpha$ appears due to relativistic corrections. 

The change in a frequency ratio $\omega_1/\omega_2$ caused by a change in $\alpha$ is
\begin{equation}
\delta\left(\frac{\omega_1}{\omega_2}\right)=
\frac{\delta\omega_1}{\omega_1} - \frac{\delta\omega_2}{\omega_2} = \left(K_1 - K_2 \right)\frac{\delta\alpha}{\alpha}.
\end{equation}
The value $K=2q/\omega$ is often called the enhancement factor. We calculate $q$ and $K$ by using two different values of $\alpha$ and calculating the numerical derivative
\begin{equation}
q=\frac{\omega(x)-\omega(-x)}{2x},
\end{equation}
where $x=(\alpha/\alpha_0)^2-1$ [see Eq.~(\ref{e:q})]. In order to achieve linear behaviour, the $x$ value must be small; however, it must be large enough to suppress numerical noise. Accurate results can be obtained by using $x=0.01$. A summary of the calculated values of $q$ and $K$ is given in Table ~\ref{t:q}. 

\begin{table}[!]
	\caption{\label{t:q}
		Sensitivity of clock transitions to variation of the fine-structure constant ($q, K$) .} 
	
	\begin{ruledtabular}
		\begin{tabular}{lrccc}
			
			\multicolumn{1}{c}{System}& 
			\multicolumn{1}{c}{State}&

			\multicolumn{1}{c}{$\omega$ (cm$^{-1}$)}&
			
			\multicolumn{1}{c}{ $q$ (cm$^{-1}$) }&
				\multicolumn{1}{c}{$K$ }\\

			\hline
				\multicolumn{5}{c}{\textbf{Te-like systems}}\\

		 Te& $5p^{4}$~$^3${P}$_{0}$&4706&3261&1.39\\


		Xe$^{2+}$&$5s^{2}5p^{4}$~$^3${P}$_{0}$&8130&5611&1.38\\


	Ba$^{4+}$&$5s^{2}5p^{4}$~$^3${P}$_{0}$& 11302&5976&1.06\\


	Ce$^{6+}$&$5s^{2}5p^{4}$~$^3${P}$_{0}$&14210&5907&0.83\\

	\multicolumn{5}{c}{\textbf{Se-like systems}}\\
	
	Zr$^{6+}$&$4s^{2}4p^{4}$~$^3${P}$_{0}$&12557&8939&1.42\\
	
	Cd$^{14+}$&$4s^{2}4p^{4}$~$^3${P}$_{0}$&28828&8837&0.61\\

									\multicolumn{5}{c}{\textbf{S-like systems}}\\
						
 	Ge$^{16+}$&$3s^{2}3p^{4}$~$^3${P}$_{0}$&33290&18484&1.11\\

		Kr$^{20+}$&$3s^{2}3p^{4}$~$^3${P}$_{0}$&46900&17252&0.74\\

		Sr$^{22+}$&$3p^{4}$~$^3${P}$_{0}$&53400&14130&0.53\\	
			
		\end{tabular}
		
	\end{ruledtabular}
	
\end{table}

\section{\label{sec:Expt}Experimental outlook}
Here, we discuss the experimental outlook for the development of optical atomic clocks based on these systems.  The systematic shifts considered in previous sections are limited by the atomic properties of the respective system.  However, when estimating the expected clock performance, it is important to also consider systematic shifts due to ion motion (time dilation) and the expected frequency instability.  To estimate the frequency instability, we consider a Ramsey interrogation sequence for a single ion with interrogation time equal to the natural lifetime, assuming the instability to be limited by fundamental quantum projection noise~\cite{Itano1993PRA}.  Under these conditions, the fractional instability is given by~\cite{Peik2005JPhysB,Kozlov2018RMP}
\begin{equation}
\sigma_y(t)=\frac{0.412}{\nu\sqrt{\tau t}},
\end{equation}
where $\nu$ is the clock frequency, $\tau$ is the lifetime of the excited clock state, and $t$ is the averaging time.  These results are summarized in Table~\ref{tab:Expt}.  All systems exhibit frequency instabilities, for a single clock ion, of $\sigma_y(t) < 5 \times 10^{-16} / \sqrt{t/\rm{s}}$.  This level of performance is comparable to recent demonstrations in Al$^{+}$ and Yb$^{+}$~\cite{Brewer2019PRL,Clements2020PRL,Sanner2019Nature}.

\begin{table}[tb]
	\caption{\label{tab:Expt}
Lifetime-limited frequency instability for a single clock ion and the optimal logic ion based on charge-to-mass ratio (q/m); m are the average values of mass over all isotopes, taken from NIST data ~\cite{NIST_ASD}.
The notation $x[y]$ abbreviates $x\times10^y$.} 
	
	\begin{ruledtabular}
		\begin{tabular}{lcccc}
			
			\multicolumn{1}{c}{System}& 

			\multicolumn{1}{c}{$\nu$ [Hz]}&
			
			\multicolumn{1}{c}{ $\sigma_{y}$ (1s) }&
				\multicolumn{1}{c}{ Logic Ion}&
		\multicolumn{1}{c}{$\dfrac{(\rm q/m) _{ Logic \ Ion}}{(\rm q/m) _{ Clock\ Ion}}$}\\

	\hline
	\multicolumn{5}{c}{\textbf{Te-like systems}}\\

	Te & 1.411[14] & $2.6[-16]$ & $ - $ &$ - $\\
	
	
	Xe$^{2+}$ & 2.437[14] & $3.4[-16]$ & Sr$^{+}$&0.749\\
	
	
	Ba$^{4+}$ & 3.388[14] & $4.1[-16]$ & Ca$^{+}$&0.857\\
	
	
	Ce$^{6+}$ & 4.260[14] & $4.5[-16]$ & Mg$^{+}$&0.961\\

	\multicolumn{5}{c}{\textbf{Se-like systems}}\\
	
	Zr$^{6+}$ & 3.764[14] & $2.3[-16]$ & Be$^{+}$&1.687\\
	
	Cd$^{14+}$ & 8.642[14] & $4.2[-16]$ & Be$^{+}$&0.891\\

	\multicolumn{5}{c}{\textbf{S-like systems}}\\
	
	Ge$^{16+}$ & 9.980[14] & $2.0[-16]$ & Be$^{+}$&0.504\\
	
	Kr$^{20+}$ & 1.406[15] & $2.6[-16]$ & Be$^{+}$&0.465\\
	
	Sr$^{22+}$ & 1.601[15] & $2.9[-16]$ & Be$^{+}$&0.442\\	
	
	\end{tabular}
	
	\end{ruledtabular}
	
\end{table}
Since none of the ions proposed here possess electric dipole-allowed ($E1$) transitions for cooling and state readout, it will be necessary to utilize a scheme such as quantum-logic spectroscopy (QLS) for clock operations~\cite{Schmidt2005Science}.  The application of QLS requires the clock ion to be co-trapped with an auxiliary readout ``logic" ion which does possess a laser-accessible transition for cooling and state readout operations.  In addition, ion-based optical clocks are susceptible to time-dilation shifts due to driven excess micromotion (EMM) and secular (thermal) motion due to the finite ion temperature.  The secular motion can be reduced by applying sympathetic cooling of the clock ion via the co-trapped logic ion.  The most efficient sympathetic cooling occurs when the charge-to-mass ratio of the clock ion is equal to that of the logic ion~\cite{Wubbena2012PRA}.  For each ion considered here, we estimate the logic ion which would be the best match for sympathetic cooling.  These results are listed in Table~\ref{tab:Expt}.

The excess micromotion shift is a result of imperfections in the trap potential, typically caused by stray electric fields and/or phase shifts between rf drive electrodes that lead to residual rf fields at the location of the ion~\cite{Berkeland1998JAP}.  This shift can be minimized by using a trap design which has been shown to have low EMM~\cite{Brewer2019PRL, Pyka2013APB}.

\section{Summary}
In conclusion, we identify group 16-like ions as promising candidates for high-accuracy optical clocks.  This class of ions exhibit irregular ordering in the ground $^3P_J$ fine structure manifold for large enough ionization degree, leading to $E2$ clock transitions with narrow natural linewidths.  Due to the increased charge state, several common systematic shifts are reduced compared to many of the current species used for optical clocks.

\begin{acknowledgments}
The authors thanks C.-C.~Chen, G.~Hoth, and D.~Slichter for their careful reading of the manuscript S. O. Allehabi gratefully acknowledges the Islamic University of Madinah (Ministry of Education, Kingdom of Saudi Arabia) for funding his scholarship. This work was supported by the National Institute of Standards and Technology/Physical Measurement Laboratory. This work was supported by the Australian Research Council Grants No. DP190100974 and DP200100150 and by NSF Grant No.~PHY-2110102 and ONR Grant No.~N00014-22-1-2070.. This research includes computations using the computational cluster Katana supported by Research Technology Services at UNSW Sydney~\cite{Katana}.
\end{acknowledgments}

\bibliography{referencesFinal}

\begin{thebibliography}{47}%
\makeatletter
\providecommand \@ifxundefined [1]{%
 \@ifx{#1\undefined}
}%
\providecommand \@ifnum [1]{%
 \ifnum #1\expandafter \@firstoftwo
 \else \expandafter \@secondoftwo
 \fi
}%
\providecommand \@ifx [1]{%
 \ifx #1\expandafter \@firstoftwo
 \else \expandafter \@secondoftwo
 \fi
}%
\providecommand \natexlab [1]{#1}%
\providecommand \enquote  [1]{``#1''}%
\providecommand \bibnamefont  [1]{#1}%
\providecommand \bibfnamefont [1]{#1}%
\providecommand \citenamefont [1]{#1}%
\providecommand \href@noop [0]{\@secondoftwo}%
\providecommand \href [0]{\begingroup \@sanitize@url \@href}%
\providecommand \@href[1]{\@@startlink{#1}\@@href}%
\providecommand \@@href[1]{\endgroup#1\@@endlink}%
\providecommand \@sanitize@url [0]{\catcode `\\12\catcode `\$12\catcode
  `\&12\catcode `\#12\catcode `\^12\catcode `\_12\catcode `\%12\relax}%
\providecommand \@@startlink[1]{}%
\providecommand \@@endlink[0]{}%
\providecommand \url  [0]{\begingroup\@sanitize@url \@url }%
\providecommand \@url [1]{\endgroup\@href {#1}{\urlprefix }}%
\providecommand \urlprefix  [0]{URL }%
\providecommand \Eprint [0]{\href }%
\providecommand \doibase [0]{https://doi.org/}%
\providecommand \selectlanguage [0]{\@gobble}%
\providecommand \bibinfo  [0]{\@secondoftwo}%
\providecommand \bibfield  [0]{\@secondoftwo}%
\providecommand \translation [1]{[#1]}%
\providecommand \BibitemOpen [0]{}%
\providecommand \bibitemStop [0]{}%
\providecommand \bibitemNoStop [0]{.\EOS\space}%
\providecommand \EOS [0]{\spacefactor3000\relax}%
\providecommand \BibitemShut  [1]{\csname bibitem#1\endcsname}%
\let\auto@bib@innerbib\@empty
\bibitem [{\citenamefont {Ludlow}\ \emph {et~al.}(2015)\citenamefont {Ludlow}
  \emph {et~al.}}]{Ludlow2015RMP}%
  \BibitemOpen
  \bibfield  {author} {\bibinfo {author} {\bibfnamefont {A.~D.}\ \bibnamefont
  {Ludlow}} \emph {et~al.},\ }\bibfield  {title} {\bibinfo {title} {Optical
  atomic clocks},\ }\href@noop {} {\bibfield  {journal} {\bibinfo  {journal}
  {Rev. Mod. Phys.}\ }\textbf {\bibinfo {volume} {87}},\ \bibinfo {pages} {637}
  (\bibinfo {year} {2015})}\BibitemShut {NoStop}%
\bibitem [{\citenamefont {Safronova}\ \emph {et~al.}(2018)\citenamefont
  {Safronova} \emph {et~al.}}]{Safronova2018BSM}%
  \BibitemOpen
  \bibfield  {author} {\bibinfo {author} {\bibfnamefont {M.~S.}\ \bibnamefont
  {Safronova}} \emph {et~al.},\ }\bibfield  {title} {\bibinfo {title} {Search
  for new physics with atoms and molecules},\ }\href
  {https://doi.org/10.1103/RevModPhys.90.025008} {\bibfield  {journal}
  {\bibinfo  {journal} {Rev. Mod. Phys.}\ }\textbf {\bibinfo {volume} {90}},\
  \bibinfo {pages} {025008} (\bibinfo {year} {2018})}\BibitemShut {NoStop}%
\bibitem [{\citenamefont {Brewer}\ \emph {et~al.}(2019)\citenamefont {Brewer}
  \emph {et~al.}}]{Brewer2019PRL}%
  \BibitemOpen
  \bibfield  {author} {\bibinfo {author} {\bibfnamefont {S.~M.}\ \bibnamefont
  {Brewer}} \emph {et~al.},\ }\href@noop {} {\bibfield  {journal} {\bibinfo
  {journal} {Phys. Rev. Lett.}\ }\textbf {\bibinfo {volume} {123}},\ \bibinfo
  {pages} {033201} (\bibinfo {year} {2019})}\BibitemShut {NoStop}%
\bibitem [{\citenamefont {McGrew}\ \emph {et~al.}(2018)\citenamefont {McGrew}
  \emph {et~al.}}]{McGrew2018Nature}%
  \BibitemOpen
  \bibfield  {author} {\bibinfo {author} {\bibfnamefont {W.~F.}\ \bibnamefont
  {McGrew}} \emph {et~al.},\ }\bibfield  {title} {\bibinfo {title} {Atomic
  clock performance enabling geodesy below the centimetre level},\ }\href
  {https://doi.org/10.1038/s41586-018-0738-2} {\bibfield  {journal} {\bibinfo
  {journal} {Nature}\ }\textbf {\bibinfo {volume} {564}},\ \bibinfo {pages}
  {87} (\bibinfo {year} {2018})}\BibitemShut {NoStop}%
\bibitem [{\citenamefont {Bothwell}\ \emph {et~al.}(2019)\citenamefont
  {Bothwell}, \citenamefont {Kedar}, \citenamefont {Oelker}, \citenamefont
  {Robinson}, \citenamefont {Bromley}, \citenamefont {Tew}, \citenamefont
  {Ye},\ and\ \citenamefont {Kennedy}}]{Bothwell2019Metrologia}%
  \BibitemOpen
  \bibfield  {author} {\bibinfo {author} {\bibfnamefont {T.}~\bibnamefont
  {Bothwell}}, \bibinfo {author} {\bibfnamefont {D.}~\bibnamefont {Kedar}},
  \bibinfo {author} {\bibfnamefont {E.}~\bibnamefont {Oelker}}, \bibinfo
  {author} {\bibfnamefont {J.~M.}\ \bibnamefont {Robinson}}, \bibinfo {author}
  {\bibfnamefont {S.~L.}\ \bibnamefont {Bromley}}, \bibinfo {author}
  {\bibfnamefont {W.~L.}\ \bibnamefont {Tew}}, \bibinfo {author} {\bibfnamefont
  {J.}~\bibnamefont {Ye}},\ and\ \bibinfo {author} {\bibfnamefont {C.~J.}\
  \bibnamefont {Kennedy}},\ }\bibfield  {title} {\bibinfo {title} {{JILA} {SrI}
  optical lattice clock with uncertainty of $2.0\times10^{-18}$},\ }\href
  {https://doi.org/10.1088/1681-7575/ab4089} {\bibfield  {journal} {\bibinfo
  {journal} {Metrologia}\ }\textbf {\bibinfo {volume} {56}},\ \bibinfo {pages}
  {065004} (\bibinfo {year} {2019})}\BibitemShut {NoStop}%
\bibitem [{\citenamefont {Huntemann}\ \emph {et~al.}(2016)\citenamefont
  {Huntemann} \emph {et~al.}}]{Huntemann2016PRL}%
  \BibitemOpen
  \bibfield  {author} {\bibinfo {author} {\bibfnamefont {N.}~\bibnamefont
  {Huntemann}} \emph {et~al.},\ }\bibfield  {title} {\bibinfo {title}
  {Single-ion atomic clock with $3 \times 10^{-18}$ systematic uncertainty},\
  }\href@noop {} {\bibfield  {journal} {\bibinfo  {journal} {Phys. Rev. Lett.}\
  }\textbf {\bibinfo {volume} {116}},\ \bibinfo {pages} {063001} (\bibinfo
  {year} {2016})}\BibitemShut {NoStop}%
\bibitem [{\citenamefont {Berengut}\ \emph {et~al.}(2010)\citenamefont
  {Berengut}, \citenamefont {Dzuba},\ and\ \citenamefont {Flambaum}}]{HCI1}%
  \BibitemOpen
  \bibfield  {author} {\bibinfo {author} {\bibfnamefont {J.~C.}\ \bibnamefont
  {Berengut}}, \bibinfo {author} {\bibfnamefont {V.~A.}\ \bibnamefont
  {Dzuba}},\ and\ \bibinfo {author} {\bibfnamefont {V.~V.}\ \bibnamefont
  {Flambaum}},\ }\bibfield  {title} {\bibinfo {title} {Enhanced laboratory
  sensitivity to variation of the fine structure constant using highly-charged
  ions},\ }\href@noop {} {\bibfield  {journal} {\bibinfo  {journal} {Phys. Rev.
  Lett.}\ }\textbf {\bibinfo {volume} {105}},\ \bibinfo {pages} {120801}
  (\bibinfo {year} {2010})}\BibitemShut {NoStop}%
\bibitem [{\citenamefont {Berengut}\ \emph {et~al.}(2011)\citenamefont
  {Berengut}, \citenamefont {Dzuba}, \citenamefont {Flambaum},\ and\
  \citenamefont {Ong}}]{HCI2}%
  \BibitemOpen
  \bibfield  {author} {\bibinfo {author} {\bibfnamefont {J.~C.}\ \bibnamefont
  {Berengut}}, \bibinfo {author} {\bibfnamefont {V.~A.}\ \bibnamefont {Dzuba}},
  \bibinfo {author} {\bibfnamefont {V.~V.}\ \bibnamefont {Flambaum}},\ and\
  \bibinfo {author} {\bibfnamefont {A.}~\bibnamefont {Ong}},\ }\bibfield
  {title} {\bibinfo {title} {Electron-hole transitions in multiply-charged ions
  for precision laser spectroscopy and searching for alpha-variation},\
  }\href@noop {} {\bibfield  {journal} {\bibinfo  {journal} {Phys. Rev. Lett.}\
  }\textbf {\bibinfo {volume} {106}},\ \bibinfo {pages} {210802} (\bibinfo
  {year} {2011})}\BibitemShut {NoStop}%
\bibitem [{\citenamefont {Berengut}\ \emph
  {et~al.}(2012{\natexlab{a}})\citenamefont {Berengut}, \citenamefont {Dzuba},
  \citenamefont {Flambaum},\ and\ \citenamefont {Ong}}]{HCI3}%
  \BibitemOpen
  \bibfield  {author} {\bibinfo {author} {\bibfnamefont {J.~C.}\ \bibnamefont
  {Berengut}}, \bibinfo {author} {\bibfnamefont {V.~A.}\ \bibnamefont {Dzuba}},
  \bibinfo {author} {\bibfnamefont {V.~V.}\ \bibnamefont {Flambaum}},\ and\
  \bibinfo {author} {\bibfnamefont {A.}~\bibnamefont {Ong}},\ }\bibfield
  {title} {\bibinfo {title} {Optical transitions in highly-charged californium
  ions with high sensitivity to variation of the fine-structure constant},\
  }\href@noop {} {\bibfield  {journal} {\bibinfo  {journal} {Phys. Rev. Lett.}\
  }\textbf {\bibinfo {volume} {109}},\ \bibinfo {pages} {070802} (\bibinfo
  {year} {2012}{\natexlab{a}})}\BibitemShut {NoStop}%
\bibitem [{\citenamefont {Kozlov}\ \emph {et~al.}(2018)\citenamefont {Kozlov},
  \citenamefont {Safronova}, \citenamefont {Crespo L\'opez-Urrutia},\ and\
  \citenamefont {Schmidt}}]{Kozlov2018RMP}%
  \BibitemOpen
  \bibfield  {author} {\bibinfo {author} {\bibfnamefont {M.~G.}\ \bibnamefont
  {Kozlov}}, \bibinfo {author} {\bibfnamefont {M.~S.}\ \bibnamefont
  {Safronova}}, \bibinfo {author} {\bibfnamefont {J.~R.}\ \bibnamefont {Crespo
  L\'opez-Urrutia}},\ and\ \bibinfo {author} {\bibfnamefont {P.~O.}\
  \bibnamefont {Schmidt}},\ }\bibfield  {title} {\bibinfo {title} {Highly
  charged ions: Optical clocks and applications in fundamental physics},\
  }\href {https://doi.org/10.1103/RevModPhys.90.045005} {\bibfield  {journal}
  {\bibinfo  {journal} {Rev. Mod. Phys.}\ }\textbf {\bibinfo {volume} {90}},\
  \bibinfo {pages} {045005} (\bibinfo {year} {2018})}\BibitemShut {NoStop}%
\bibitem [{\citenamefont {Berengut}\ \emph
  {et~al.}(2012{\natexlab{b}})\citenamefont {Berengut}, \citenamefont {Dzuba},
  \citenamefont {Flambaum},\ and\ \citenamefont {Ong}}]{HCI4}%
  \BibitemOpen
  \bibfield  {author} {\bibinfo {author} {\bibfnamefont {J.~C.}\ \bibnamefont
  {Berengut}}, \bibinfo {author} {\bibfnamefont {V.~A.}\ \bibnamefont {Dzuba}},
  \bibinfo {author} {\bibfnamefont {V.~V.}\ \bibnamefont {Flambaum}},\ and\
  \bibinfo {author} {\bibfnamefont {A.}~\bibnamefont {Ong}},\ }\bibfield
  {title} {\bibinfo {title} {Highly charged ions with e1, m1, and e2
  transitions within laser range},\ }\href@noop {} {\bibfield  {journal}
  {\bibinfo  {journal} {Phys. Rev. A}\ }\textbf {\bibinfo {volume} {86}},\
  \bibinfo {pages} {022517} (\bibinfo {year} {2012}{\natexlab{b}})}\BibitemShut
  {NoStop}%
\bibitem [{\citenamefont {Kramida}\ \emph {et~al.}(2020)\citenamefont
  {Kramida}, \citenamefont {Ralchenko}, \citenamefont {Reader},\ and\
  \citenamefont {Team}}]{NIST_ASD}%
  \BibitemOpen
  \bibfield  {author} {\bibinfo {author} {\bibfnamefont {A.}~\bibnamefont
  {Kramida}}, \bibinfo {author} {\bibfnamefont {Y.}~\bibnamefont {Ralchenko}},
  \bibinfo {author} {\bibfnamefont {J.}~\bibnamefont {Reader}},\ and\ \bibinfo
  {author} {\bibfnamefont {N.~A.}\ \bibnamefont {Team}},\ }\href@noop {} {}
  (\bibinfo {year} {2020}),\ \bibinfo {note} {{NIST Atomic Spectra Database
  (ver. 5.7.1), [Online]. Available: https://physics.nist.gov/asd [2020,
  September 26]. National Institute of Standards and Technology, Gaithersburg,
  MD. DOI: https://doi.org/10.18434/T4W30F}}\BibitemShut {NoStop}%
\bibitem [{\citenamefont {Cheng}\ \emph {et~al.}(1979)\citenamefont {Cheng},
  \citenamefont {Kim},\ and\ \citenamefont {Desclaux}}]{CHENG1979ADNDT}%
  \BibitemOpen
  \bibfield  {author} {\bibinfo {author} {\bibfnamefont {K.}~\bibnamefont
  {Cheng}}, \bibinfo {author} {\bibfnamefont {Y.-K.}\ \bibnamefont {Kim}},\
  and\ \bibinfo {author} {\bibfnamefont {J.}~\bibnamefont {Desclaux}},\
  }\bibfield  {title} {\bibinfo {title} {Electric dipole, quadrupole, and
  magnetic dipole transition probabilities of ions isoelectronic to the
  first-row atoms, li through f},\ }\href
  {https://doi.org/https://doi.org/10.1016/0092-640X(79)90006-8} {\bibfield
  {journal} {\bibinfo  {journal} {Atomic Data and Nuclear Data Tables}\
  }\textbf {\bibinfo {volume} {24}},\ \bibinfo {pages} {111} (\bibinfo {year}
  {1979})}\BibitemShut {NoStop}%
\bibitem [{\citenamefont {Gayasov}\ \emph {et~al.}(1997)\citenamefont
  {Gayasov}, \citenamefont {Joshi},\ and\ \citenamefont
  {Tauheed}}]{Gayasov1997JPhsB}%
  \BibitemOpen
  \bibfield  {author} {\bibinfo {author} {\bibfnamefont {R.}~\bibnamefont
  {Gayasov}}, \bibinfo {author} {\bibfnamefont {Y.~N.}\ \bibnamefont {Joshi}},\
  and\ \bibinfo {author} {\bibfnamefont {A.}~\bibnamefont {Tauheed}},\
  }\bibfield  {title} {\bibinfo {title} {Sixth spectrum of lanthanum (la vi):
  analysis of the 5s$^2$5p$^4$, 5s5p$^5$ and 5s$^2$5p$^3$(5d + 6s)
  configurations},\ }\href@noop {} {\bibfield  {journal} {\bibinfo  {journal}
  {J. Phys. B: At. Mol. Opt. Phys.}\ }\textbf {\bibinfo {volume} {30}},\
  \bibinfo {pages} {873} (\bibinfo {year} {1997})}\BibitemShut {NoStop}%
\bibitem [{\citenamefont {Tauheed}\ and\ \citenamefont
  {Joshi}(2008)}]{Tauheed2008CanJPhys}%
  \BibitemOpen
  \bibfield  {author} {\bibinfo {author} {\bibfnamefont {A.}~\bibnamefont
  {Tauheed}}\ and\ \bibinfo {author} {\bibfnamefont {Y.~N.}\ \bibnamefont
  {Joshi}},\ }\bibfield  {title} {\bibinfo {title} {The 5s$^2$5p$^4$-(5s5p$^5$
  + 5p$^3$6s ) transitions in ce vii and 5s$^2$5p$^3$ $^4$s - 5s5p$^4$ $^4p$
  transitions in ce viii},\ }\href@noop {} {\bibfield  {journal} {\bibinfo
  {journal} {Can. J. Phys.}\ }\textbf {\bibinfo {volume} {86}},\ \bibinfo
  {pages} {713} (\bibinfo {year} {2008})}\BibitemShut {NoStop}%
\bibitem [{\citenamefont {Reader}\ and\ \citenamefont
  {Acquista}(1976)}]{Reader1976JOSA}%
  \BibitemOpen
  \bibfield  {author} {\bibinfo {author} {\bibfnamefont {J.}~\bibnamefont
  {Reader}}\ and\ \bibinfo {author} {\bibfnamefont {N.}~\bibnamefont
  {Acquista}},\ }\bibfield  {title} {\bibinfo {title} {4s$^2$4p$^4$-4s4p$^5$
  transitions in zr {\small vii}, nb viii, and mo ix},\ }\href@noop {}
  {\bibfield  {journal} {\bibinfo  {journal} {J. Opt. Soc. Am.}\ }\textbf
  {\bibinfo {volume} {66}},\ \bibinfo {pages} {896} (\bibinfo {year}
  {1976})}\BibitemShut {NoStop}%
\bibitem [{\citenamefont {Beloy}\ \emph {et~al.}(2020)\citenamefont {Beloy},
  \citenamefont {Dzuba},\ and\ \citenamefont {Brewer}}]{Beloy2020PRL}%
  \BibitemOpen
  \bibfield  {author} {\bibinfo {author} {\bibfnamefont {K.}~\bibnamefont
  {Beloy}}, \bibinfo {author} {\bibfnamefont {V.~A.}\ \bibnamefont {Dzuba}},\
  and\ \bibinfo {author} {\bibfnamefont {S.~M.}\ \bibnamefont {Brewer}},\
  }\bibfield  {title} {\bibinfo {title} {Quadruply ionized barium as a
  candidate for a high-accuracy optical clock},\ }\href
  {https://doi.org/10.1103/PhysRevLett.125.173002} {\bibfield  {journal}
  {\bibinfo  {journal} {Phys. Rev. Lett.}\ }\textbf {\bibinfo {volume} {125}},\
  \bibinfo {pages} {173002} (\bibinfo {year} {2020})}\BibitemShut {NoStop}%
\bibitem [{\citenamefont {Berkeland}\ \emph {et~al.}(1998)\citenamefont
  {Berkeland} \emph {et~al.}}]{Berkeland1998JAP}%
  \BibitemOpen
  \bibfield  {author} {\bibinfo {author} {\bibfnamefont {D.~J.}\ \bibnamefont
  {Berkeland}} \emph {et~al.},\ }\bibfield  {title} {\bibinfo {title}
  {Minimization of ion micromotion in a paul trap},\ }\href@noop {} {\bibfield
  {journal} {\bibinfo  {journal} {J. Appl. Phys.}\ }\textbf {\bibinfo {volume}
  {83}},\ \bibinfo {pages} {5025} (\bibinfo {year} {1998})}\BibitemShut
  {NoStop}%
\bibitem [{\citenamefont {Dzuba}(2014)}]{Dzuba2014PRA}%
  \BibitemOpen
  \bibfield  {author} {\bibinfo {author} {\bibfnamefont {V.~A.}\ \bibnamefont
  {Dzuba}},\ }\bibfield  {title} {\bibinfo {title} {Combination of the
  single-double--coupled-cluster and the configuration-interaction methods:
  Application to barium, lutetium, and their ions},\ }\href
  {https://doi.org/10.1103/PhysRevA.90.012517} {\bibfield  {journal} {\bibinfo
  {journal} {Phys. Rev. A}\ }\textbf {\bibinfo {volume} {90}},\ \bibinfo
  {pages} {012517} (\bibinfo {year} {2014})}\BibitemShut {NoStop}%
\bibitem [{\citenamefont {Johnson}\ and\ \citenamefont
  {Sapirstein}(1986)}]{Johnson1986PRL}%
  \BibitemOpen
  \bibfield  {author} {\bibinfo {author} {\bibfnamefont {W.~R.}\ \bibnamefont
  {Johnson}}\ and\ \bibinfo {author} {\bibfnamefont {J.}~\bibnamefont
  {Sapirstein}},\ }\bibfield  {title} {\bibinfo {title} {Computation of
  second-order many-body corrections in relativistic atomic systems},\ }\href
  {https://doi.org/10.1103/PhysRevLett.57.1126} {\bibfield  {journal} {\bibinfo
   {journal} {Phys. Rev. Lett.}\ }\textbf {\bibinfo {volume} {57}},\ \bibinfo
  {pages} {1126} (\bibinfo {year} {1986})}\BibitemShut {NoStop}%
\bibitem [{\citenamefont {Johnson}\ \emph {et~al.}(1988)\citenamefont
  {Johnson}, \citenamefont {Blundell},\ and\ \citenamefont
  {Sapirstein}}]{Johnson1988PRA}%
  \BibitemOpen
  \bibfield  {author} {\bibinfo {author} {\bibfnamefont {W.~R.}\ \bibnamefont
  {Johnson}}, \bibinfo {author} {\bibfnamefont {S.~A.}\ \bibnamefont
  {Blundell}},\ and\ \bibinfo {author} {\bibfnamefont {J.}~\bibnamefont
  {Sapirstein}},\ }\bibfield  {title} {\bibinfo {title} {Finite basis sets for
  the dirac equation constructed from b splines},\ }\href
  {https://doi.org/10.1103/PhysRevA.37.307} {\bibfield  {journal} {\bibinfo
  {journal} {Phys. Rev. A}\ }\textbf {\bibinfo {volume} {37}},\ \bibinfo
  {pages} {307} (\bibinfo {year} {1988})}\BibitemShut {NoStop}%
\bibitem [{\citenamefont {Dzuba}\ \emph
  {et~al.}(1987{\natexlab{a}})\citenamefont {Dzuba}, \citenamefont {Flambaum},
  \citenamefont {Silvestrov},\ and\ \citenamefont {Sushkov}}]{CPM}%
  \BibitemOpen
  \bibfield  {author} {\bibinfo {author} {\bibfnamefont {V.~A.}\ \bibnamefont
  {Dzuba}}, \bibinfo {author} {\bibfnamefont {V.~V.}\ \bibnamefont {Flambaum}},
  \bibinfo {author} {\bibfnamefont {P.~G.}\ \bibnamefont {Silvestrov}},\ and\
  \bibinfo {author} {\bibfnamefont {O.~P.}\ \bibnamefont {Sushkov}},\
  }\bibfield  {title} {\bibinfo {title} {Correlation potential method for the
  calculation of energy levels, hyperfine structure and e1 transition
  amplitudes in atoms with one unpaired electron},\ }\href@noop {} {\bibfield
  {journal} {\bibinfo  {journal} {J. Phys. B: {\it At. Mol. Phys.}}\ }\textbf
  {\bibinfo {volume} {20}},\ \bibinfo {pages} {1399} (\bibinfo {year}
  {1987}{\natexlab{a}})}\BibitemShut {NoStop}%
\bibitem [{\citenamefont {Dzuba}\ \emph {et~al.}(1996)\citenamefont {Dzuba},
  \citenamefont {Flambaum},\ and\ \citenamefont {Kozlov}}]{Sigma2}%
  \BibitemOpen
  \bibfield  {author} {\bibinfo {author} {\bibfnamefont {V.~A.}\ \bibnamefont
  {Dzuba}}, \bibinfo {author} {\bibfnamefont {V.~V.}\ \bibnamefont
  {Flambaum}},\ and\ \bibinfo {author} {\bibfnamefont {M.~G.}\ \bibnamefont
  {Kozlov}},\ }\bibfield  {title} {\bibinfo {title} {Combination of the
  many-body perturbation theory with the configuration interaction method},\
  }\href@noop {} {\bibfield  {journal} {\bibinfo  {journal} {Phys. Rev. A}\
  }\textbf {\bibinfo {volume} {54}},\ \bibinfo {pages} {3948} (\bibinfo {year}
  {1996})}\BibitemShut {NoStop}%
\bibitem [{\citenamefont {Dzuba}\ \emph {et~al.}(2017)\citenamefont {Dzuba},
  \citenamefont {Berengut}, \citenamefont {Harabati},\ and\ \citenamefont
  {Flambaum}}]{Dzuba2017PRA}%
  \BibitemOpen
  \bibfield  {author} {\bibinfo {author} {\bibfnamefont {V.~A.}\ \bibnamefont
  {Dzuba}}, \bibinfo {author} {\bibfnamefont {J.~C.}\ \bibnamefont {Berengut}},
  \bibinfo {author} {\bibfnamefont {C.}~\bibnamefont {Harabati}},\ and\
  \bibinfo {author} {\bibfnamefont {V.~V.}\ \bibnamefont {Flambaum}},\
  }\bibfield  {title} {\bibinfo {title} {Combining configuration interaction
  with perturbation theory for atoms with a large number of valence
  electrons},\ }\href {https://doi.org/10.1103/PhysRevA.95.012503} {\bibfield
  {journal} {\bibinfo  {journal} {Phys. Rev. A}\ }\textbf {\bibinfo {volume}
  {95}},\ \bibinfo {pages} {012503} (\bibinfo {year} {2017})}\BibitemShut
  {NoStop}%
\bibitem [{\citenamefont {Dzuba}\ \emph
  {et~al.}(1987{\natexlab{b}})\citenamefont {Dzuba}, \citenamefont {Flambaum},
  \citenamefont {Silvestrov},\ and\ \citenamefont {Sushkov}}]{Dzuba1987JPhysB}%
  \BibitemOpen
  \bibfield  {author} {\bibinfo {author} {\bibfnamefont {V.~A.}\ \bibnamefont
  {Dzuba}}, \bibinfo {author} {\bibfnamefont {V.~V.}\ \bibnamefont {Flambaum}},
  \bibinfo {author} {\bibfnamefont {P.~G.}\ \bibnamefont {Silvestrov}},\ and\
  \bibinfo {author} {\bibfnamefont {O.~P.}\ \bibnamefont {Sushkov}},\
  }\bibfield  {title} {\bibinfo {title} {Correlation potential method for the
  calculation of energy levels, hyperfine structure and e1 transition
  amplitudes in atoms with one unpaired electron},\ }\href
  {https://doi.org/10.1088/0022-3700/20/7/009} {\bibfield  {journal} {\bibinfo
  {journal} {Journal of Physics B: Atomic and Molecular Physics}\ }\textbf
  {\bibinfo {volume} {20}},\ \bibinfo {pages} {1399} (\bibinfo {year}
  {1987}{\natexlab{b}})}\BibitemShut {NoStop}%
\bibitem [{\citenamefont {Wang}\ \emph {et~al.}(2017)\citenamefont {Wang},
  \citenamefont {Yang}, \citenamefont {Chen}, \citenamefont {Si}, \citenamefont
  {Chen}, \citenamefont {Yan}, \citenamefont {Zhao},\ and\ \citenamefont
  {Dang}}]{Wang2017ADNDT}%
  \BibitemOpen
  \bibfield  {author} {\bibinfo {author} {\bibfnamefont {K.}~\bibnamefont
  {Wang}}, \bibinfo {author} {\bibfnamefont {X.}~\bibnamefont {Yang}}, \bibinfo
  {author} {\bibfnamefont {Z.}~\bibnamefont {Chen}}, \bibinfo {author}
  {\bibfnamefont {R.}~\bibnamefont {Si}}, \bibinfo {author} {\bibfnamefont
  {C.}~\bibnamefont {Chen}}, \bibinfo {author} {\bibfnamefont {J.}~\bibnamefont
  {Yan}}, \bibinfo {author} {\bibfnamefont {X.~H.}\ \bibnamefont {Zhao}},\ and\
  \bibinfo {author} {\bibfnamefont {W.}~\bibnamefont {Dang}},\ }\bibfield
  {title} {\bibinfo {title} {Energy levels, lifetimes, and transition rates for
  the selenium isoelectronic sequence pd xiii$-$te xix, xe xxi$-$nd xxvii, w
  xli},\ }\href@noop {} {\bibfield  {journal} {\bibinfo  {journal} {At Data
  Nucl Data Tables}\ }\textbf {\bibinfo {volume} {117}},\ \bibinfo {pages} {1}
  (\bibinfo {year} {2017})}\BibitemShut {NoStop}%
\bibitem [{\citenamefont {Garstang}(1964)}]{Garstand1964JRNBSAPC}%
  \BibitemOpen
  \bibfield  {author} {\bibinfo {author} {\bibfnamefont {R.~H.}\ \bibnamefont
  {Garstang}},\ }\bibfield  {title} {\bibinfo {title} {Transition probabilities
  of forbidden lines},\ }\href@noop {} {\bibfield  {journal} {\bibinfo
  {journal} {J. Res. Natl. Bur. Stand. A Phys. Chem.}\ }\textbf {\bibinfo
  {volume} {68}},\ \bibinfo {pages} {61} (\bibinfo {year} {1964})}\BibitemShut
  {NoStop}%
\bibitem [{\citenamefont {Biémont}\ \emph {et~al.}(1995)\citenamefont
  {Biémont}, \citenamefont {Hansen}, \citenamefont {Quinet},\ and\
  \citenamefont {Zeippen}}]{Biemont1995AASS}%
  \BibitemOpen
  \bibfield  {author} {\bibinfo {author} {\bibfnamefont {E.}~\bibnamefont
  {Biémont}}, \bibinfo {author} {\bibfnamefont {J.~E.}\ \bibnamefont
  {Hansen}}, \bibinfo {author} {\bibfnamefont {P.}~\bibnamefont {Quinet}},\
  and\ \bibinfo {author} {\bibfnamefont {C.~J.}\ \bibnamefont {Zeippen}},\
  }\bibfield  {title} {\bibinfo {title} {Forbidden transitions of astrophysical
  interest in the 5p$^k$ (k= 1-5) configurations},\ }\href@noop {} {\bibfield
  {journal} {\bibinfo  {journal} {Astron. Astrophys., Suppl. Ser.}\ }\textbf
  {\bibinfo {volume} {111}},\ \bibinfo {pages} {333} (\bibinfo {year}
  {1995})}\BibitemShut {NoStop}%
\bibitem [{\citenamefont {Biémont}\ and\ \citenamefont
  {Hansen}(1986)}]{Biemont1986PhysScr}%
  \BibitemOpen
  \bibfield  {author} {\bibinfo {author} {\bibfnamefont {E.}~\bibnamefont
  {Biémont}}\ and\ \bibinfo {author} {\bibfnamefont {J.~E.}\ \bibnamefont
  {Hansen}},\ }\bibfield  {title} {\bibinfo {title} {Forbidden transitions in
  3p$^4$ and 4p$^4$ configurations},\ }\href@noop {} {\bibfield  {journal}
  {\bibinfo  {journal} {Phys. Scr.}\ }\textbf {\bibinfo {volume} {34}},\
  \bibinfo {pages} {116} (\bibinfo {year} {1986})}\BibitemShut {NoStop}%
\bibitem [{\citenamefont {Schwerdtfegera}\ and\ \citenamefont
  {Nagle}(2019)}]{Schwerdtfegera2019MolPhys}%
  \BibitemOpen
  \bibfield  {author} {\bibinfo {author} {\bibfnamefont {P.}~\bibnamefont
  {Schwerdtfegera}}\ and\ \bibinfo {author} {\bibfnamefont {J.~K.}\
  \bibnamefont {Nagle}},\ }\bibfield  {title} {\bibinfo {title} {2018 table of
  static dipole polarizabilities of theneutral elements in the periodic
  table},\ }\href@noop {} {\bibfield  {journal} {\bibinfo  {journal} {Molecular
  Physics}\ }\textbf {\bibinfo {volume} {117}},\ \bibinfo {pages} {1200}
  (\bibinfo {year} {2019})}\BibitemShut {NoStop}%
\bibitem [{\citenamefont {Itano}(2000)}]{Itano2000NISTJRes}%
  \BibitemOpen
  \bibfield  {author} {\bibinfo {author} {\bibfnamefont {W.}~\bibnamefont
  {Itano}},\ }\bibfield  {title} {\bibinfo {title} {External-field shifts of
  the $^{199}{\rm hg}^+$ optical frequency standard},\ }\href@noop {}
  {\bibfield  {journal} {\bibinfo  {journal} {J. Res. Natl. Inst. Stand.
  Technol.}\ }\textbf {\bibinfo {volume} {105}},\ \bibinfo {pages} {829}
  (\bibinfo {year} {2000})}\BibitemShut {NoStop}%
\bibitem [{\citenamefont {Kozlov}\ \emph {et~al.}(2014)\citenamefont {Kozlov},
  \citenamefont {Dzuba},\ and\ \citenamefont {Flambaum}}]{Kozlov2014PRA}%
  \BibitemOpen
  \bibfield  {author} {\bibinfo {author} {\bibfnamefont {A.}~\bibnamefont
  {Kozlov}}, \bibinfo {author} {\bibfnamefont {V.~A.}\ \bibnamefont {Dzuba}},\
  and\ \bibinfo {author} {\bibfnamefont {V.~V.}\ \bibnamefont {Flambaum}},\
  }\bibfield  {title} {\bibinfo {title} {Optical atomic clocks with suppressed
  blackbody-radiation shift},\ }\href
  {https://doi.org/10.1103/PhysRevA.90.042505} {\bibfield  {journal} {\bibinfo
  {journal} {Phys. Rev. A}\ }\textbf {\bibinfo {volume} {90}},\ \bibinfo
  {pages} {042505} (\bibinfo {year} {2014})}\BibitemShut {NoStop}%
\bibitem [{\citenamefont {Dzuba}(2020)}]{Dzuba2020Symmetry}%
  \BibitemOpen
  \bibfield  {author} {\bibinfo {author} {\bibfnamefont {V.}~\bibnamefont
  {Dzuba}},\ }\bibfield  {title} {\bibinfo {title} {Calculation of
  polarizabilities for atoms with open shells},\ }\href@noop {} {\bibfield
  {journal} {\bibinfo  {journal} {Symmetry}\ }\textbf {\bibinfo {volume}
  {12}},\ \bibinfo {pages} {1950} (\bibinfo {year} {2020})}\BibitemShut
  {NoStop}%
\bibitem [{\citenamefont {Dalgarno}\ and\ \citenamefont
  {Lewis}(1955)}]{Dalgarno1955PRS}%
  \BibitemOpen
  \bibfield  {author} {\bibinfo {author} {\bibfnamefont {A.}~\bibnamefont
  {Dalgarno}}\ and\ \bibinfo {author} {\bibfnamefont {J.~T.}\ \bibnamefont
  {Lewis}},\ }\bibfield  {title} {\bibinfo {title} {The exact calculation of
  longrange forces between atoms by perturbation theory},\ }\href@noop {}
  {\bibfield  {journal} {\bibinfo  {journal} {Proc. R. Soc. London A}\ }\textbf
  {\bibinfo {volume} {233}},\ \bibinfo {pages} {70} (\bibinfo {year}
  {1955})}\BibitemShut {NoStop}%
\bibitem [{\citenamefont {Dzuba}\ \emph {et~al.}(2002)\citenamefont {Dzuba},
  \citenamefont {Flambaum}, \citenamefont {Ginges},\ and\ \citenamefont
  {Kozlov}}]{EDM6}%
  \BibitemOpen
  \bibfield  {author} {\bibinfo {author} {\bibfnamefont {V.~A.}\ \bibnamefont
  {Dzuba}}, \bibinfo {author} {\bibfnamefont {V.~V.}\ \bibnamefont {Flambaum}},
  \bibinfo {author} {\bibfnamefont {J.~S.~M.}\ \bibnamefont {Ginges}},\ and\
  \bibinfo {author} {\bibfnamefont {M.~G.}\ \bibnamefont {Kozlov}},\ }\bibfield
   {title} {\bibinfo {title} {Electric dipole moments of hg, xe, rn, ra, pu,
  and tlf induced by the nuclear schiff moment and limits on time-reversal
  violating interactions},\ }\href@noop {} {\bibfield  {journal} {\bibinfo
  {journal} {Phys. Rev. A}\ }\textbf {\bibinfo {volume} {66}},\ \bibinfo
  {pages} {012111} (\bibinfo {year} {2002})}\BibitemShut {NoStop}%
\bibitem [{\citenamefont {Dzuba}\ and\ \citenamefont
  {Flambaum}(2009)}]{EDM-Tl}%
  \BibitemOpen
  \bibfield  {author} {\bibinfo {author} {\bibfnamefont {V.~A.}\ \bibnamefont
  {Dzuba}}\ and\ \bibinfo {author} {\bibfnamefont {V.~V.}\ \bibnamefont
  {Flambaum}},\ }\bibfield  {title} {\bibinfo {title} {Calculation of the (t ,
  p)-odd electric dipole moment of thallium and cesium},\ }\href@noop {}
  {\bibfield  {journal} {\bibinfo  {journal} {Phys. Rev. A}\ }\textbf {\bibinfo
  {volume} {80}},\ \bibinfo {pages} {062509} (\bibinfo {year}
  {2009})}\BibitemShut {NoStop}%
\bibitem [{\citenamefont {Dzuba}\ and\ \citenamefont
  {Derevianko}(2010)}]{DD-Yb10}%
  \BibitemOpen
  \bibfield  {author} {\bibinfo {author} {\bibfnamefont {V.~A.}\ \bibnamefont
  {Dzuba}}\ and\ \bibinfo {author} {\bibfnamefont {A.}~\bibnamefont
  {Derevianko}},\ }\bibfield  {title} {\bibinfo {title} {Dynamic
  polarizabilities and related properties of clock states of the ytterbium
  atom},\ }\href@noop {} {\bibfield  {journal} {\bibinfo  {journal} {J. Phys.
  B}\ }\textbf {\bibinfo {volume} {43}},\ \bibinfo {pages} {074011} (\bibinfo
  {year} {2010})}\BibitemShut {NoStop}%
\bibitem [{\citenamefont {Porsev}\ and\ \citenamefont
  {Safronova}(2020)}]{Porsev2020PRA}%
  \BibitemOpen
  \bibfield  {author} {\bibinfo {author} {\bibfnamefont {S.~G.}\ \bibnamefont
  {Porsev}}\ and\ \bibinfo {author} {\bibfnamefont {M.~S.}\ \bibnamefont
  {Safronova}},\ }\bibfield  {title} {\bibinfo {title} {Calculation of
  higher-order corrections to the light shift of the $5{s}^{2}$
  $^{1}{S}_{0}\ensuremath{-}5s5p$ $^{3}{P}_{0}^{o}$ clock transition in cd},\
  }\href {https://doi.org/10.1103/PhysRevA.102.012811} {\bibfield  {journal}
  {\bibinfo  {journal} {Phys. Rev. A}\ }\textbf {\bibinfo {volume} {102}},\
  \bibinfo {pages} {012811} (\bibinfo {year} {2020})}\BibitemShut {NoStop}%
\bibitem [{\citenamefont {Flambaum}\ and\ \citenamefont
  {Dzuba}(2009)}]{Flambaum2009CanJPhys}%
  \BibitemOpen
  \bibfield  {author} {\bibinfo {author} {\bibfnamefont {V.~V.}\ \bibnamefont
  {Flambaum}}\ and\ \bibinfo {author} {\bibfnamefont {V.~A.}\ \bibnamefont
  {Dzuba}},\ }\bibfield  {title} {\bibinfo {title} {Search for variation of the
  fundamental constants in atomic, molecular, and nuclear spectra},\
  }\href@noop {} {\bibfield  {journal} {\bibinfo  {journal} {Can. J. Phys.}\
  }\textbf {\bibinfo {volume} {87}},\ \bibinfo {pages} {25} (\bibinfo {year}
  {2009})}\BibitemShut {NoStop}%
\bibitem [{\citenamefont {Itano}\ \emph {et~al.}(1993)\citenamefont {Itano}
  \emph {et~al.}}]{Itano1993PRA}%
  \BibitemOpen
  \bibfield  {author} {\bibinfo {author} {\bibfnamefont {W.~M.}\ \bibnamefont
  {Itano}} \emph {et~al.},\ }\bibfield  {title} {\bibinfo {title} {Quantum
  projection noise: Pupulation flucturation in two-level systems},\ }\href@noop
  {} {\bibfield  {journal} {\bibinfo  {journal} {Phys. Rev. A}\ }\textbf
  {\bibinfo {volume} {47}},\ \bibinfo {pages} {3554} (\bibinfo {year}
  {1993})}\BibitemShut {NoStop}%
\bibitem [{\citenamefont {Peik}\ \emph {et~al.}(2005)\citenamefont {Peik},
  \citenamefont {Schneider},\ and\ \citenamefont {Tamm}}]{Peik2005JPhysB}%
  \BibitemOpen
  \bibfield  {author} {\bibinfo {author} {\bibfnamefont {E.}~\bibnamefont
  {Peik}}, \bibinfo {author} {\bibfnamefont {T.}~\bibnamefont {Schneider}},\
  and\ \bibinfo {author} {\bibfnamefont {C.}~\bibnamefont {Tamm}},\ }\bibfield
  {title} {\bibinfo {title} {Laser frequency stabilization to a single ion},\
  }\href {https://doi.org/10.1088/0953-4075/39/1/012} {\bibfield  {journal}
  {\bibinfo  {journal} {J. Phys. B}\ }\textbf {\bibinfo {volume} {39}},\
  \bibinfo {pages} {145} (\bibinfo {year} {2005})}\BibitemShut {NoStop}%
\bibitem [{\citenamefont {Clements}\ \emph {et~al.}(2020)\citenamefont
  {Clements} \emph {et~al.}}]{Clements2020PRL}%
  \BibitemOpen
  \bibfield  {author} {\bibinfo {author} {\bibfnamefont {E.~R.}\ \bibnamefont
  {Clements}} \emph {et~al.},\ }\bibfield  {title} {\bibinfo {title}
  {Lifetime-limited interrogation of two independent $^{27}{\mathrm{al}}^{+}$
  clocks using correlation spectroscopy},\ }\href
  {https://doi.org/10.1103/PhysRevLett.125.243602} {\bibfield  {journal}
  {\bibinfo  {journal} {Phys. Rev. Lett.}\ }\textbf {\bibinfo {volume} {125}},\
  \bibinfo {pages} {243602} (\bibinfo {year} {2020})}\BibitemShut {NoStop}%
\bibitem [{\citenamefont {Sanner}\ \emph {et~al.}(2019)\citenamefont {Sanner},
  \citenamefont {Huntemann}, \citenamefont {Lange}, \citenamefont {Tamm},
  \citenamefont {Peik}, \citenamefont {Safronova},\ and\ \citenamefont
  {Porsev}}]{Sanner2019Nature}%
  \BibitemOpen
  \bibfield  {author} {\bibinfo {author} {\bibfnamefont {C.}~\bibnamefont
  {Sanner}}, \bibinfo {author} {\bibfnamefont {N.}~\bibnamefont {Huntemann}},
  \bibinfo {author} {\bibfnamefont {R.}~\bibnamefont {Lange}}, \bibinfo
  {author} {\bibfnamefont {C.}~\bibnamefont {Tamm}}, \bibinfo {author}
  {\bibfnamefont {E.}~\bibnamefont {Peik}}, \bibinfo {author} {\bibfnamefont
  {M.~S.}\ \bibnamefont {Safronova}},\ and\ \bibinfo {author} {\bibfnamefont
  {S.~G.}\ \bibnamefont {Porsev}},\ }\bibfield  {title} {\bibinfo {title}
  {Optical clock comparison for lorentz symmetry testing},\ }\href@noop {}
  {\bibfield  {journal} {\bibinfo  {journal} {Nature}\ }\textbf {\bibinfo
  {volume} {567}},\ \bibinfo {pages} {204} (\bibinfo {year}
  {2019})}\BibitemShut {NoStop}%
\bibitem [{\citenamefont {Schmidt}\ \emph {et~al.}(2005)\citenamefont {Schmidt}
  \emph {et~al.}}]{Schmidt2005Science}%
  \BibitemOpen
  \bibfield  {author} {\bibinfo {author} {\bibfnamefont {P.~O.}\ \bibnamefont
  {Schmidt}} \emph {et~al.},\ }\bibfield  {title} {\bibinfo {title}
  {Spectroscopy using quantum logic},\ }\href@noop {} {\bibfield  {journal}
  {\bibinfo  {journal} {Science}\ }\textbf {\bibinfo {volume} {309}},\ \bibinfo
  {pages} {749} (\bibinfo {year} {2005})}\BibitemShut {NoStop}%
\bibitem [{\citenamefont {W\"ubbena}\ \emph {et~al.}(2012)\citenamefont
  {W\"ubbena}, \citenamefont {Amairi}, \citenamefont {Mandel},\ and\
  \citenamefont {Schmidt}}]{Wubbena2012PRA}%
  \BibitemOpen
  \bibfield  {author} {\bibinfo {author} {\bibfnamefont {J.~B.}\ \bibnamefont
  {W\"ubbena}}, \bibinfo {author} {\bibfnamefont {S.}~\bibnamefont {Amairi}},
  \bibinfo {author} {\bibfnamefont {O.}~\bibnamefont {Mandel}},\ and\ \bibinfo
  {author} {\bibfnamefont {P.~O.}\ \bibnamefont {Schmidt}},\ }\bibfield
  {title} {\bibinfo {title} {Sympathetic cooling of mixed-species two-ion
  crystals for precision spectroscopy},\ }\href
  {https://doi.org/10.1103/PhysRevA.85.043412} {\bibfield  {journal} {\bibinfo
  {journal} {Phys. Rev. A}\ }\textbf {\bibinfo {volume} {85}},\ \bibinfo
  {pages} {043412} (\bibinfo {year} {2012})}\BibitemShut {NoStop}%
\bibitem [{\citenamefont {Pyka}\ \emph {et~al.}(2013)\citenamefont {Pyka},
  \citenamefont {Herschbach}, \citenamefont {Keller},\ and\ \citenamefont
  {Mehlst\"aubler}}]{Pyka2013APB}%
  \BibitemOpen
  \bibfield  {author} {\bibinfo {author} {\bibfnamefont {K.}~\bibnamefont
  {Pyka}}, \bibinfo {author} {\bibfnamefont {N.}~\bibnamefont {Herschbach}},
  \bibinfo {author} {\bibfnamefont {J.}~\bibnamefont {Keller}},\ and\ \bibinfo
  {author} {\bibfnamefont {T.~E.}\ \bibnamefont {Mehlst\"aubler}},\ }\bibfield
  {title} {\bibinfo {title} {A high-precision segmented paul trap with
  minimized micromotioin for an optical multiple-ion clock},\ }\href@noop {}
  {\bibfield  {journal} {\bibinfo  {journal} {Appl. Phys. B}\ }\textbf
  {\bibinfo {volume} {114}},\ \bibinfo {pages} {231} (\bibinfo {year}
  {2013})}\BibitemShut {NoStop}%
\bibitem [{Kat(2010)}]{Katana}%
  \BibitemOpen
  \href@noop {} {} (\bibinfo {year} {2010}),\ \bibinfo {note} {katana (shared
  computational cluster), [https://doi.org/10.26190/669x-a286] University of
  New South Wales, Sydney}\BibitemShut {NoStop}%
\end{thebibliography}%

\end{document}